\documentclass{svjour3}                     
\smartqed  
\usepackage{graphicx}
\journalname{Few-Body Systems}
\usepackage{graphics}
\usepackage{epsfig}
\usepackage{textpos,colortbl,xcolor}
\usepackage{fancybox,fancyhdr}
\usepackage{amsmath,amssymb,bm}
\usepackage{enumitem}
\newcommand{\be}{\begin{equation}}
\newcommand{\ee}{\end{equation}}

\newcommand{\ber}{\begin{eqnarray}}
\newcommand{\eer}{\end{eqnarray}}

\newcommand{\bra}{\langle}
\newcommand{\ket}{\rangle}

\newcommand{\bs}[1]{\ensuremath{\boldsymbol{#1}}}
\newcommand{\rvec}{\ensuremath{\boldsymbol{r}}}
\newcommand{\Rvec}{\ensuremath{\boldsymbol{R}}}

\newcommand{\qvec}{\ensuremath{\boldsymbol{q}}}
\newcommand{\kvec}{\ensuremath{\boldsymbol{k}}}
\newcommand{\Kvec}{\ensuremath{\boldsymbol{K}}}
\begin{document}

\title{Short range correlations - The important role of few-body dynamics in
many-body systems
}


\author{Ronen Weiss \and Ehoud Pazy \and Nir Barnea }   


\institute{The Racah Institute of Physics, The Hebrew University, \at
              91904 Givaat-Ram, Jerusalem, Israel \\
              \email{nir@phys.huji.ac.il}           
           \and
           Department of Physics, \at NRCN, P.O.B. 9001, \at Beer-Sheva 84190, Israel
}

\date{Received: date / Accepted: date}

\maketitle
\begin{abstract}
For many-body systems with short range interaction a series of relations
were derived connecting many properties of the system to the dynamics of
a closely packed few-body subsystems.
Some of these relations were experimentally verified in ultra cold atomic 
gases.
Here we shall review the implications of these developments on our understanding of
nuclear one and two-body momentum distributions, and on the electron scattering 
Coulomb sum rule.

\end{abstract}
\section{Introduction}
Recent advances in the study of many-body systems suggest that the behavior of
systems composed of particles interacting via short-range force  
 are governed by the probability of finding a particle pair or
triplet in a close proximity. 
Considering a system of two-component fermions interacting via ``zero-range'' 
$s$-wave forces, Tan \cite{Tan08} and later others 
(see {\it e.g.} \cite{Bra12,WerCas12} and references therein)
have established a series 
of relations between the amplitude of the high-momentum tail of the momentum distribution 
$n(k)$ and many properties of the system. 
These relations, commonly known as the ``Tan relations'',
are expressed through a new state variable, the ``Contact'' $C$, 
that for the aforementioned system dominates the tail of the momentum 
distribution 
$C=\lim_{k\rightarrow\infty}k^4 n(k)$. The contact $C$ is a measure 
for the probability of finding a 
particle pair close to each other, or in other words,
a measure
for the short range correlations (SRCs) in the system.
The Tan relations are universal, they hold for any few-body 
or many-body system where 
the interparticle distance $d$,
and the magnitude of the scattering length $a_s$
are both much larger than the potential range $R_{pot}$, {\it i.e.} $a_s,d \gg R_{pot}$,
and the average particle momentum $k$ is much smaller than $1/R_{pot}$.

Experiments done in two-component ultra-cold atomic Fermi gases, such as  
 $^{40}$K \cite{SteGaeDra10,SagDraPau12} and $^6$Li 
\cite{ParStrKam05,WerTarCas09,KuhHuLiu10},
have tested the Tan relations.
Using different experimental techniques
and comparing several observables with the predictions of the universal theory,
the Tan relations were verified.

Following Tan's seminal work on two-component fermions, similar relations 
were derived for one-component fermions, and for bosons.
Due to the Pauli principle, in a system of one-component fermions, 
the low energy interaction is predominantly $p$-wave. In this case,  
the momentum distribution $n(k)$ falls as $1/k^2$ for large momentum
and the correponding contact is given by
$C=\lim_{k\rightarrow\infty}k^2 n(k)$ \cite{nature-pwave-2016}. 

The Tan relations for bosons interacting via zero-range force, have much in common
with a system of two-component fermions. The main difference is the emergence
of important 
 3-body correlations associated with the 3-body force 
\cite{3BodyTerm}, 
that appear in low energy effective theory 
(For a review, see {\it e.g.} Ref. \cite{pionlesseft})
to prevent the Thomas collapse \cite{Thomas35}.
In this case the high momentum tail of the momentum distribution aquires a 
$1/k^5$ component modulated 
 by a log-periodic function associated with the Efimov effect \cite{Efi70}.

The atomic nuclei exhibit some key properties similar to these 
idealized systems. They are made of fermions, the nuclear force is short range,
and the scattering length is much larger than the potential range.
Therefore one expects SRCs to play an important role in nuclear physics
(for literature see for example 
 \cite{Lev51,CioSim96,Ciofi15,HenSci14,AlvCio13,AlvCioMor08,ArrHigRos12,Wiringa14,Fomin12}), and the
notion of the contact to be a useful concept in this context.

To apply Tan's ideas to nuclear systems we need to note also the differences 
between the atomic nuclei and the aforementioned systems.
The nucleon can be regarded as a four-component fermion 
$N=(p{\mkern-5mu}\uparrow,p{\mkern-5mu}\downarrow,
  n{\mkern-6mu}\uparrow,n{\mkern-6mu}\downarrow)$, 
the protons are charged particles interacting via
the long range Coulomb force,
and most importantly 
the range of the nuclear force (dictated by the pion mass $m_\pi$)
 in not much shorter than the average interparticle 
distance, therefore the ``zero range'' 
approximation doesn't really hold. 

The similarities between atomic nuclei and ``zero-range'' fermionic or bosonic
systems suggest that the
contact formalism can be a good starting point to study nuclear systems. 
The differences suggest that Tan's ideas must be generalized 
to accommodate nuclear systems, and that different or modified
``relations'' are expected to hold.
For example, there is no obvious reason to assume that the asymptotic tail of the nuclear
momentum distribution will take on Tan's form  $\lim_{k\rightarrow\infty}n(k)=C/k^4$.  

In the last couple of years the utility of
the contact formalism in nuclear physics was started to be explored.
The neutron-proton $s$-wave nuclear contacts have been defined and evaluated 
\cite{WeissPRL15,Hen15}, 
relating their value to the Levinger photoabsorption constant \cite{Lev51,TavTer92},
and to high energy inclusive electron scattering \cite{Hen15,Ciofi16}. 
Considering not only $s$-wave but
all partial waves, as well as finite-range interactions instead
of zero-range, it was found that the nuclear SRCs are governed 
by a set of contact matrices \cite{WeissPRC15}. 
Using this generalized contact formalism to relate
the nuclear contacts to the one-nucleon and two-nucleon momentum distributions,
an asymptotic relation between these two 
distributions was established \cite{WeissPRC15}, which is relevant to the study of SRCs in nuclei. 
This relation was verified using available numerical data \cite{Wiringa14}. 
The contact formalism was further applied to study the nuclear equation of state and symmetry
energy (see {\it e.g.} \cite{Bao-Jun16} and references therein).

In this paper we first review Tan's relations for one-body and two-body
momentum distributions,
and their generalization to different ideal systems, Sec. \ref{sec_single_channel}. 
Then in Sec. \ref{sec_nuclear_contacts} we
present the generalization of these relations to nuclear systems \cite{WeissPRC15}. 
Finally we utilize this formalism to derive the asymptotic behavior of the electron
scattering Coulomb sum rule, Sec. \ref{sec_csr}. 

Given our limited understanding of the role of 3-body effects in nuclear SRCs,
throughout this paper we focus on the nuclear two-body contacts, and their applications.

\section{Tan's contact - the single channel case}
\label{sec_single_channel}

Consider an $N$--particle system that obeys Tan's assumptions
({\it i.e.} $a_s,d \gg R_{pot}$, and $k \ll 1/R_{pot}$), 
and is dominated by a single
interaction channel such as an $s$-wave interaction in low energy Bose gas, or
$p$-wave in a single-component Fermi gas.
In such a gas, when interacting particle pair $(ij)$ get close together, 
the many-body wave function $\Psi$ can be factorized into a 
product of an asymptotic pair wave function $\varphi(\bs r_{ij})$,
$\bs r_{ij}=\bs{r}_i-\bs{r}_j$,
and a function $A$, also called the regular part of $\Psi$,
describing the residual $N-2$ particle system
and the pair's center of mass $\bs{R}_{ij}=(\bs{r}_i+\bs{r}_j)/2$
motion \cite{Tan08,WerCas12},
\be \label{wf}
  \Psi \xrightarrow[r_{ij}\rightarrow 0]{}\varphi(\bs{r}_{ij})
           A(\bs{R}_{ij},\{\bs r_k\}_{k\neq i,j})\;.
\ee
The asymptotic pair wave function $\varphi$ depends on the interparticle potential
and the dominant channel. In particular, in the zero-range model \cite{zerorange} 
the $s$-wave function is given by $\varphi=\left(1/r_{ij}-1/a_s\right)$, where $a_s$ is
the scattering length.
The contact $C$ is then defined by \cite{Tan08,WerCas12}
\be\label{contact_generic}
   C=16\pi^2 N_{pairs} \bra A| A\ket,
\ee
where 
\begin{align}
   \bra A|  A\ket & = 
    \int  \prod_{k\neq i,j} d\bs{r}_{k} \,d\bs{R}_{ij} \,
        A^{\dagger}\left(\bs{R}_{ij},\{\bs{r}_{k}\}_{k\neq i,j}\right)
        \cdot
        A\left(\bs{R}_{ij},\{\bs{r}_{k}\}_{k\neq i,j}\right)\;
\end{align}
and $N_{pairs}$ is the number of interacting pairs. For a Bose or single-component Fermi gas
$N_{pairs}=N(N-1)/2$, for two-component fermions $N_{pairs}=N_{\uparrow}N_{\downarrow}$.

Working in the momentum space,
\be
   \tilde{\Psi}(\bs{k}_1,...,\bs{k}_A)=\int  
      \!\prod_{n=1}^A d^3\rvec_n\;  e^{i\sum_{n}\bs{k}_n\cdot\bs{r}_n}\;\Psi(\rvec_1,\ldots ,\rvec_A),
\ee
the short range factorization takes on the form of high momentum factorization.
When a particle pair $ij$ approach each other $r_{ij}\longrightarrow 0$ 
the relative momentum $\kvec_{ij}=(\kvec_i-\kvec_j)/2$ diverges $k_{ij}\longrightarrow\infty$.
In this limit
\be \label{psi_tilde}
  \tilde\Psi \xrightarrow[k_{ij}\rightarrow \infty]{}\tilde\varphi(\kvec_{ij})
           \tilde{A}(\Kvec_{ij},\{\kvec_n\}_{n\neq i,j})\;,
\ee
where $\Kvec_{ij}=\kvec_i+\kvec_j$, $\tilde\varphi$ is the Fourier transform of $\varphi$,
and $\tilde A$ is the Fourier transform of $A$.

Using these definitions it is easy to see that the contact can be equivalently written 
as
\be 
   C=16\pi^2 N_{pairs} \bra \tilde A| \tilde A\ket, 
\ee
since 
\begin{align} \label{atilde_norm}
   \bra \tilde A| \tilde A\ket 
       &= 
       \int \!\frac{d\Kvec_{ij}}{(2\pi)^3} \!\prod_{n\neq i,j} \!\frac{d \kvec_n}{(2\pi)^3} 
        \tilde{A}^{\dagger}\left(\Kvec_{ij},\{\kvec_{n}\}_{n\neq i,j}\right)
        \cdot
        \tilde{A}\left(\Kvec_{ij},\{\kvec_{n}\}_{n\neq i,j}\right)
      \cr & = \bra A|  A\ket. 
\end{align}

For finite systems such as nuclei and for non homogeneous infinite matter
it can be important to consider two-body 
center of mass (CM) effects. Therefore it is
convenient to introduce the {\it  contact function} \cite{Tan08}
\be\label{C_of_K}
  C(\Kvec_{ij}) = 16 \pi^2 N_{pairs}
        \int \!\!\prod_{n\neq i,j} \!\frac{d \kvec_n}{(2\pi)^3} 
        \tilde{A}^{\dagger}\left(\Kvec_{ij},\{\kvec_{n}\}_{n\neq i,j}\right)
        \cdot
        \tilde{A}\left(\Kvec_{ij},\{\kvec_{n}\}_{n\neq i,j}\right).
\ee
An equivalent expression can also be derived for $C(\Rvec_{ij})$.
Comparing eqs. (\ref{atilde_norm}) and (\ref{C_of_K}) we see that the relation 
between the contact and the cotact function
is given through the integral
\be
  C = \int \!\frac{d\Kvec}{(2\pi)^3} C(\Kvec).
\ee

In the following sections we will present the application of these definitions to derive
Tan's relations for the one and two-body momentum distributions.
For simplicity we first start with the two--body case.

\subsection{The two-nucleon momentum distribution}
\label{sec_2b_momentum}

Consider now a pair of particles $ij$ with relative momentum $\kvec$ and CM momentum $\Kvec$.
We denote by $f(\bs{k}+\bs{K}/2, -\bs{k}+\bs{K}/2)$ the
density probability to find 
particle  $i$ with momentum $\kvec_i=\bs{k}+\bs{K}/2$ and particle $j$ 
with momentum $\kvec_j=-\bs{k}+\bs{K}/2$. 
We immediately see that
\begin{align}\label{fij_tan}
 f(\kvec_i,\kvec_j)=
      N_{pairs} \int  \!\prod_{m\neq i,j} \!\frac{d^3\kvec_m}{(2\pi)^3} 
  \left| \tilde{\Psi}(\bs{k}_1,\ldots \bs{k}_A)\right|^2.
\end{align}
Here $f$ is normalized in such away that 
$\int \frac{d^3\kvec_i}{(2\pi)^3} \frac{d^3\kvec_j}{(2\pi)^3}f(\kvec_i,\kvec_j)=N_{pairs}$.

For very large relative momentum
the main contribution to $f(\kvec_i,\kvec_j)$
comes from the asymptotic $k\rightarrow \infty$ part of the wave function,
given in Eq. (\ref{psi_tilde}). 
All other terms will vanish
due to the 
fast oscillating $\exp(i\bs{k}\cdot \bs{r}_{ij})$ factor.
Using Eq. (\ref{psi_tilde}) and substituting $\tilde{\Psi}$ into Eq. (\ref{fij_tan}), 
we get
\be\label{fij_asymp_tan}
   f(\bs{k}+\bs{K}/2, -\bs{k}+\bs{K}/2)= \frac{C(\Kvec)}{16\pi^2}
       |\tilde{\varphi}(\kvec)|^2  \;.
\ee
Thus we see that in the limit $k\longrightarrow \infty$ the two-body momentum 
distribution is given by a product of the contact function and the universal momentum distribution
$|\tilde{\varphi}(\kvec)|^2$. For particles interacting via zero-range $s$-wave potential
$\varphi \approx (1/r-1/a) $ and $\tilde\varphi \approx 4\pi/k^2$. Similarly for
the $p$-wave case $\tilde\varphi \approx 4\pi/k$.

The probability density to find a pair with relative momentum $\bs{k}$
is obtained by integrating over the CM momentum
\be
  F(\bs{k})=\int \frac{d\Kvec}{(2\pi)^3}  f(\bs{k}+\bs{K}/2, -\bs{k}+\bs{K}/2).
\ee
Utilizing now the relation between the contact function and the contact
we can now substitute the asymptotic form of $f$, Eq. (\ref{fij_asymp_tan}), 
 to get
\be \label{2b_momentum}
   F(\bs{k})= \frac{C}{16\pi^2}|\tilde{\varphi}(\kvec)|^2  
\ee

\subsection{The one-body momentum distribution}
\label{sec_1b_momentum}

We would like to relate now the contact also to the 
single particle momentum distribution. To this end we will follow
Tan's derivation for the two-body case \cite{Tan08}.
To simplify the notation for the moment we will only consider the case of bosons and
one-component fermions. 

Normalized to the number of particles in the system,
 $\int\frac{d\kvec}{(2\pi)^3} n(\bs{k})=N$,
the single particle momentum distribution $n(\kvec)$ is given by
\begin{align}\label{n_p_tan}
 n(\bs{k})& = N \int \prod_{l \neq p} \frac{d \kvec_l}{(2\pi)^3} 
      \left| \tilde{\Psi}(\bs{k}_{1},...,\bs{k}_{p}=\bs{k},...,\bs{k}_{A})\right|^2,
\end{align}
where $p$ is any particle.

In the $k\longrightarrow \infty$ limit the main contribution to $n(\kvec)$
emerges from the asymptotic parts of the wave function, i.e.
from $r_{p s}=|\bs{r}_{p}-\bs{r}_s|\rightarrow 0$, for any particle $s \neq p$.
In this limit
\begin{align} 
\tilde{\Psi}(\bs{k}_{1},...,&\bs{k}_{p}=\bs{k},...,\bs{k}_{A})= 
 \sum_{s \neq p} \tilde{\varphi}\left(\bs{k}-\Kvec_{ps}/2\right) 
      \tilde{A}\left(\Kvec_{p s},\{\bs{k}_j\}_{j \neq p,s}\right),
\end{align}
where $\bs{K}_{p s}= \kvec_p+\kvec_s$ is the center of mass momentum of the $p s$ pair.
We note that $\bs{k}$ is fixed in (\ref{n_p_tan}) while integrating over all other momenta.
Therefore
we can replace the integration $\int d\bs{k}_s$ with integration over the pair's
center of mass momentum $\int d\Kvec_{ps}$.
Substituting this result into Eq. (\ref{n_p_tan}), 
we get summations over 
particles $s$ and $s'$ different from $p$.
The contribution of non diagonal $s \neq s'$ terms,
will be significant only for $\bs{k}_s\approx \bs{k}_{s'} \approx -\bs{k}$,
due to the regularity of $A$.
In this case $k,k_s,k_{s'}\rightarrow \infty$ together, which
is clearly a three body effect, and we expect it to be less important 
than the leading two-body contribution \cite{BraKanPla11}.
Consequently, we only consider the diagonal elements and obtain
\begin{align}\label{1b_momentum_a}
  n(\bs{k})&= N \sum_{s \neq p} 
     \int \frac{d\Kvec_{p s}}{(2\pi)^3}\!\prod_{l \neq p, s} \frac{d\kvec_l}{(2\pi)^3}
          |\tilde{\varphi}(\kvec-\Kvec_{ps}/2) |^2 
          |\tilde{A}(\bs{K}_{p s},\{\bs{k}_j\}_{j \neq p,s}) |^2
  \cr    &=     
         2 \int \frac{d\Kvec}{(2\pi)^3}
           \frac{C(\Kvec)}{16\pi^2}
          |\tilde{\varphi}(\bs{k}-\Kvec/2)|^2
\end{align}
Deriving this result we have utilized the definition of the contact function,
Eq. (\ref{C_of_K}). The prefactor 2, results from the number of interacting
pairs that for bosons or one-component fermions is given by $N_{pairs} = N(N-1)/2$.
Consequently, for two-component fermions one obtains the same result up to this factor of 2.

Since $A$ is regular, we expect $C(\Kvec)$
to be significant only for $\Kvec_{ps}$ of the order of the average interparticle 
distance $1/d$.
Therefore $K$ can be considered to be much smaller than $k$. 
Expanding $|\tilde{\varphi}|^2$ around $\bs{k}$,
\begin{align}
|\tilde{\varphi}\left(\bs{k}-\Kvec/2\right)|^2
   &\cong
        |\tilde{\varphi}\left(\bs{k}\right)|^2 
                 -\frac{\Kvec}{2}\cdot
     \left(\tilde\varphi^\dagger(\kvec)\nabla_k\tilde{\varphi}(\kvec)+
           \nabla_k\tilde\varphi^\dagger(\kvec)\tilde{\varphi}(\kvec) \right)
 +\ldots
\end{align}
and keeping only the leading order, which is a good
approximation for any power-law function,
we obtain 
\begin{align}\label{1b_momentum}
   n(\kvec) = 2 \frac{C}{16\pi^2} |\tilde{\varphi}\left(\bs{k}\right)|^2  \;.
\end{align}
Substituting now the universal $s$-wave function $\tilde{\varphi}(\kvec)\approx 4\pi/k^2$
we obtain Tan's result for two-component Fermi gas $n(\kvec)=C/k^4$, recalling of course that
for a two-component Fermi gas the factor of 2 disappears. 

Comparing Eqs. (\ref{1b_momentum}) and Eq. (\ref{2b_momentum}),
we can see that for $k\longrightarrow \infty$ there is a simple
relation between the one-body and 
the two-body momentum distributions. For bosons or one-component fermions:
\be \label{1bto2b_1}
  n(\bs{k})=2 F(\bs{k}),
\ee
and for two-component fermions
\be \label{1bto2b_2}
  n(\bs{k})=F(\bs{k}).
\ee

\section{The nuclear contact matrices - the multi channel case}
\label{sec_nuclear_contacts}

Turning now to consider nuclear physics, we regard the nucleons
as four-component fermions, 
which are the protons and neutrons with their spin being either up or down
$(p{\mkern-5mu}\uparrow,p{\mkern-5mu}\downarrow,
  n{\mkern-6mu}\uparrow,n{\mkern-6mu}\downarrow)$.
As a result in the most simplistic model of the nuclear interaction one needs to consider
at least two contacts \cite{WeissPRL15}, and in reality one needs to consider
strong coupling between channels, such as $s$-wave
and $d$-wave mixture in the deuteron.
As a result, when extending the contact formalism to nuclear physics
one needs to consider the different interaction channels and the possible
couplings between them. Here we shall follow reference \cite{WeissPRC15}.

When a
nucleon $i$ gets close to nucleon $j$, we must abandon
the factorization ansatz and write the wave function 
as a sum of products of two-body terms $\varphi_{ij}$ and $A-2$-body 
terms $A_{ij}$, taking into account all possible channels. 
The asymptotic form of the wave function is then given by
\be \label{full_asymp}
\Psi\xrightarrow[r_{ij}\rightarrow 0]{} 
          \sum_\alpha\varphi_{ij}^\alpha\left(\bs{r}_{ij}) 
                    A_{ij}^\alpha(\bs{R}_{ij},\{\bs{r}_k\}_{k\not=i,j}\right).
\ee
We note that due to symmetry the asymptotic functions are invariant under same particle
permutations.
Therefore the index $ij$ is an indicator to the particle pair type, i.e.
proton-proton ($pp$), neutron-neutron ($nn$) or neutron-proton ($np$). 
The pair wave functions depend on the total
spin of the pair $s_2$, and its angular momentum quantum number $\ell_2$ 
(with
respect to the relative coordinate $\bs{r}_{ij}$) which
are coupled to create the total pair angular momentum $j_2$
and projection $m_2$. The quantum numbers $\left(s_2,\ell_2,j_2,m_2\right)$ define the pair's
channel. In general, the expansion (\ref{full_asymp}) may contain more then one term
per channel, however in the limit $r_{ij}\rightarrow 0$ only the leading term survives.
In short, the sum over $\alpha$ denotes a sum over
the four channel quantum numbers $\left(s_2,\ell_2,j_2,m_2\right)$. 

To ensure an $A$-body wave function with total angular momentum $J$ and projection $M$
the regular functions $A_{ij}^\alpha$ are given by
\be
A_{ij}^\alpha=\sum_{J_{A-2},M_{A-2}}\bra j_2m_2J_{A-2}M_{A-2}|JM\ket A_{ij}^{\{s_2,\ell_2,j_2\}J_{A-2},M_{A-2}}
\;,
\ee
where $J_{A-2}$ and $M_{A-2}$ are the angular momentum
quantum numbers with respect to the sum
 $\bs{J}_{A-2}+\bs{L}_{2,CM}$ of the total angular
 momentum of the residual $(A-2)$ particles $\bs{J}_{A-2}$, 
and the orbital angular momentum $\bs{L}_{2,CM}$ corresponding to $\bs{R}_{ij}$.
$A_{ij}^{\{s_2,\ell_2,j_2\}J_{A-2},M_{A-2}}$ is a set of functions
with angular momentum quantum numbers $J_{A-2}$ and $M_{A-2}$,
which depends also on the numbers $s_2,\ell_2,j_2$.
\be
\varphi_{ij}^{\alpha}\equiv\varphi_{ij}^{(\ell_2s_2)j_2m_2}
  =\phi_{ij}^{\ell_2,s_2,j_2}(r_{ij})
   [Y_{\ell_2}(\hat{r}_{ij})\otimes\chi_{s_2}]^{j_2m_2}
\;,
\ee 
where $\chi_{s_2\mu_s}$ is the two-body spin function, and 
$Y_{\ell m}$ are the spherical harmonics.

An important property of the set of asymptotic functions 
$\{\varphi_{ij}^\alpha\}$
is that they are ``universal'', in the limited sense that they
do not depend on a specific nucleus or on a specific nuclear state. 
However, they
can depend on the details of the nuclear potential and therefore cannot be related in
a simple manner to the low energy scattering parameters.

Since the $A_{ij}^\alpha$ functions are not generally orthogonal for different $\alpha$, we are led 
to define matrices of nuclear contacts in the following way \cite{WeissPRC15}:
\be\label{ave_contacts_def}
  C_{ij}^{\alpha \beta}=\frac{16{\pi}^2N_{ij}}{2J+1}\sum_M \bra A_{ij}^\alpha | A_{ij}^\beta \ket
                   =\frac{16{\pi}^2N_{ij}}{2J+1}\sum_M \bra \tilde A_{ij}^\alpha 
                               | \tilde A_{ij}^\beta \ket.
\ee
As before, $\tilde A_{ij}^\alpha$ is the Fourier transform of $ A_{ij}^\alpha$,
$ij$ stands for one of the pairs: $pp$, $nn$ or $np$, 
$N_{ij}$ is the number of $ij$ pairs, and $\alpha$ and $\beta$ are the matrix
 indices. Since the magnetic quantum number $M$ is usually unknown in experiments, it is useful 
to define the averaged nuclear contacts.

In a similar fashion we can generalize the {\it contact function} (Eq. (\ref{C_of_K}))
and define the {\it contact matrix function} taking into account
CM effects  
\be\label{nuclear_C_of_K}
  C_{ij}^{\alpha \beta}(\Kvec_{ij}) = \frac{16{\pi}^2N_{ij}}{2J+1}\sum_M
        \int \!\!\prod_{n\neq i,j} \!\frac{d \kvec_n}{(2\pi)^3} 
        \tilde{A}_{ij}^{\alpha\dagger}\left(\Kvec_{ij},\{\kvec_{n}\}_{n\neq i,j}\right)
        \cdot
        \tilde{A}_{ij}^{\beta}\left(\Kvec_{ij},\{\kvec_{n}\}_{n\neq i,j}\right).
\ee

In general the matrices $C_{ij}^{\alpha\beta}$ are built from $2\times 2$ blocks,
except for the two $1\times 1$ blocks associated with the $j_2=0$ case.
Each block has a well defined $j_2,m_2$ values.
A detailed discussion of the structure of the matrices $C_{ij}^{\alpha\beta}$ is given in 
\cite{WeissPRC15}.

It should be mentioned that the factorization of the
wave function given in Eq. (\ref{full_asymp}) was used before in the
study of nuclear SRCs, see {\it e.g.} \cite{Ciofi15} and references therein. 
In these works the relation between the asymptotic many-body wave function 
and the deuteron wave function was utilized, and the corresponding contact
was defined, see {\it e.g.} \cite{CioSim96} Eq. (29). However, it was assumed that
the nuclear contact is a single number and the general structure of
nuclear contact 
matrix was not defined or analyzed.
In the following we will demonstrate the utility of the general contact formalism,
deriving analytic relations between the nuclear one-body and two-body momentum
 distributions.

\subsection{The nuclear two-nucleon momentum distribution}

Now we can utilize the generalized contact formalism  
to find a relation between the two-nucleon momentum distribution 
and the nuclear contacts.
To this end we define $f_{ij}^{JM}(\kvec_i,\kvec_j)$ to be the
two-body momentum distribution of the $ij$ pair
associated with a nuclear state $\Psi$ with magnetic
projection $M$, and $f_{ij}(\kvec_i,\kvec_j)=1/(2J+1)\sum_M f_{ij}^{JM}(\kvec_i,\kvec_j)$
to be the averaged two-body momentum distribution.

Following the footsteps and the arguments presented in Sec. \ref{sec_2b_momentum} 
we can immediately get for $k\longrightarrow\infty$
\be \label{2nuc}
f_{ij}(\kvec+\Kvec/2,\kvec-\Kvec/2)=\sum_{\alpha,\beta}
      \frac{C_{ij}^{\alpha\beta}(\Kvec)}{16\pi^2}
      \tilde{\varphi}_{ij}^{\alpha\dagger}(\kvec)\tilde{\varphi}_{ij}^\beta(\kvec).
\ee
Accordingly, we see that if
\be
   F_{ij}(\bs{k})=\int \frac{d^3K}{(2\pi)^3}  f_{ij}(\bs{k}+\bs{K}/2, -\bs{k}+\bs{K}/2),
\ee
then asymptotically \cite{WeissPRC15}
\be \label{2nuc}
F_{ij}(\bs{k})=\sum_{\alpha,\beta}
      \frac{C_{ij}^{\alpha\beta}}{16\pi^2}
      \tilde{\varphi}_{ij}^{\alpha\dagger}(\bs{k})\tilde{\varphi}_{ij}^\beta(\bs{k}).
\ee

 \subsection{The nuclear one-nucleon momentum distribution}

In a similar way we can adress the asymptotic one-nucleon momentum distribution.
We consider first the proton's momentum distribution $n_p(\bs{k})$.
For convenience we will define $n_p(\bs{k})=\sum_M n_p^{JM}(\bs{k})/(2J+1)$ to be the 
magnetic projection averaged proton momentum distribution. 
Normalized to the number of protons in the system $Z$,
 $\int\frac{d^3k}{(2\pi)^3}n_p^{JM}(\bs{k})=Z$,
$n_p^{JM}$ is given by
\begin{align}\label{n_p_jm}
  n_p^{JM}(\bs{k})&= Z \int \prod_{l \neq p} \frac{d\kvec_l}{(2\pi)^3} 
     \left| \tilde{\Psi}(\bs{k}_{1},...,\bs{k}_{p}=\bs{k},...,\bs{k}_{A})\right|^2,
\end{align}
where $p$ is any proton.

Following the arguments leading to Eq. (\ref{1b_momentum_a})
we see that
\begin{align}
  n_p^{JM}(\kvec)&= Z \sum_{s \neq p} \sum_{\alpha,\beta} 
     \int \prod_{l \neq p, s} \frac{d\kvec_l}{(2\pi)^3}\frac{d\Kvec_{p s}}{(2\pi)^3}
    \tilde{\varphi}_{p s}^{\alpha\dagger}(\kvec-\Kvec_{ps}) 
    \tilde{\varphi}_{p s}^\beta(\kvec-\Kvec_{ps}) \cr
& \times
   \tilde{A}_{p s}^{\alpha\dagger}(\Kvec_{p s},\{\kvec_j\}_{j \neq p,s}) 
   \tilde{A}_{p s}^\beta(\Kvec_{p s},\{\kvec_j\}_{j \neq p,s}).
\end{align}
We will now divide the sum 
$\sum_{s \neq p}$ into a sum over protons and a sum
over neutrons $\sum_{p' \neq p}+\sum_n$, and average over $M$. Since the asymptotic
functions $A_{pp'}^\alpha$ and $\varphi_{pp'}^\alpha$ are the same for all $pp'$ 
pairs we can take them out of the sum.
The same holds for the $np$ pairs. As a result we get
\begin{align}
    n_p(\kvec)&=2\sum_{\alpha,\beta} 
                      \int \frac{d\Kvec}{(2\pi)^3}  
                      \frac{C_{pp}^{\alpha\beta}(\Kvec)}{16\pi^2}
                     \tilde{\varphi}_{pp}^{\alpha\dagger}(\kvec-\Kvec/2) 
                                      \tilde{\varphi}_{pp}^\beta(\kvec-\Kvec/2)
\cr & +
                     \sum_{\alpha,\beta} 
                      \int \frac{d\Kvec}{(2\pi)^3}  
                      \frac{C_{pn}^{\alpha\beta}(\Kvec)}{16\pi^2}
                     \tilde{\varphi}_{pn}^{\alpha\dagger}(\kvec-\Kvec/2) 
                                      \tilde{\varphi}_{pn}^\beta(\kvec-\Kvec/2)
\end{align}
Expanding $\tilde{\varphi}_{pn}^{\alpha}(\kvec-\Kvec/2) $ around $\kvec$ and keeping the
leading term we obtain
\begin{align} \label{1p}
n_p(\bs{k})&=2\sum_{\alpha,\beta} 
    \frac{C_{pp}^{\alpha\beta}}{16\pi^2}
    \tilde{\varphi}_{pp}^{\alpha\dagger}(\bs{k}) \tilde{\varphi}_{pp}^\beta(\bs{k}) 
  \nonumber \\
& 
   +\sum_{\alpha,\beta} 
   \frac{C_{pn}^{\alpha\beta}}{16\pi^2}
   \tilde{\varphi}_{pn}^{\alpha\dagger}(\bs{k}) \tilde{\varphi}_{pn}^\beta(\bs{k}).
\end{align}
Similarly, for the neutrons:
\begin{align} \label{1n}
    n_n(\bs{k})&=2 \sum_{\alpha,\beta} 
            \frac{C_{nn}^{\alpha\beta}}{16\pi^2}
            \tilde{\varphi}_{nn}^{\alpha\dagger}(\bs{k}) \tilde{\varphi}_{nn}^\beta(\bs{k}) 
 \cr & 
                +\sum_{\alpha,\beta} 
                 \frac{C_{pn}^{\alpha\beta}}{16\pi^2}
                 \tilde{\varphi}_{pn}^{\alpha\dagger}(\bs{k}) \tilde{\varphi}_{pn}^\beta(\bs{k}).
\end{align}
Comparing Eqs. (\ref{1p}) and (\ref{1n}) to Eq. (\ref{2nuc}),
we can see that for $k\longrightarrow \infty$ there is a simple
relation between the one-nucleon and 
the two-nucleon momentum distributions \cite{WeissPRC15}:
\be \label{1pto2}
   n_p(\bs{k})=2F_{pp}(\bs{k})+F_{pn}(\bs{k})
\ee
\be \label{1nto2}
  n_n(\bs{k})=2F_{nn}(\bs{k})+F_{pn}(\bs{k}).
\ee
These connections are the nuclear analog of Eqs. (\ref{1bto2b_1})
and (\ref{1bto2b_2}).
Their validity was verified in \cite{WeissPRC15}
utilizing the numerical data of Ref. \cite{Wiringa14}.

\section{The Coulomb sum rule}
\label{sec_csr}
Sum rules are useful tools in many fields of physics.
In nuclear physics, they typically involve an integration of the response function, 
associated with transitions between the ground state and excited states 
due to an external  probe, over the spectrum with a weight function 
composed of integer powers of the energy. 
The Coulomb sum rule (CSR) is the integral over the inelastic part of the longitudinal 
electron scattering nuclear response function. The CSR dates back to 1931 when Heisenberg
studied the photoabsorption cross-section of x-rays with momentum $\qvec$ by atoms.
Its formulation for electron scattering is due to Drell and Schwartz \cite{Drell1958},
and McVoy and Van Hove \cite{McVoy1962}.

Assuming point-like particles, the CSR can be expressed as \cite{GiusySR}
\be\label{CSRdef}
   CSR \equiv \bra \Psi | \hat\rho_c^\dagger(\qvec)\hat\rho_c(\qvec) | \Psi \ket
              - \left|\bra \Psi | \hat\rho_c(\qvec) | \Psi \ket   \right|^2,
\ee
where
\be\label{rhoq}
   \hat\rho_c(\qvec) = \sum_{j=1}^Z e^{i\qvec\cdot\rvec_j}
\ee
is the Fourier transform of the charge density operator
$\hat\rho_c(\rvec)=\sum_{j=1}^Z\delta(\rvec-\rvec_j)$. The sum in Eq. (\ref{rhoq})
is understood to include protons only as essentially the CSR is a measure of  charge 
fluctuations in the nucleus.
Substituting the charge density operator (\ref{rhoq}) into (\ref{CSRdef}), one gets
\be\label{CSRME}
   \bra \Psi | \hat\rho_c^\dagger(\qvec)\hat\rho_c(\qvec) | \Psi \ket
    = Z + \bra \Psi | \sum_{i\neq j}e^{i\qvec\cdot(\rvec_i-\rvec_j)} | \Psi \ket ,
\ee
where the first contribution on the right-hand side (rhs) comes from the $i=j$ term in the sum.
Due to the rapidly oscillating exponent, in the $q \longrightarrow \infty$ limit 
 the matrix element on the rhs will be dominated by the behavior
of the wave-function when two protons approach each other.
In this limit we can replace the wave-function by its asymptotic form
(\ref{full_asymp}) and obtain
\begin{align}
   \bra \Psi | \hat\rho_c^\dagger(\qvec)\hat\rho_c(\qvec) | \Psi \ket &
    = Z + \sum_{i\neq j}\sum_{\alpha\beta} \int d\rvec_{ij}d\Rvec_{ij}\prod_{k\neq i,j}d\rvec_k
         \varphi_{pp}^{\alpha\dagger}(\rvec_{ij}) e^{i\qvec\cdot\rvec_{ij}} \varphi_{pp}^{\beta}(\rvec_{ij})
\cr & \hspace{6em} \times 
         A^{\alpha\dagger}_{pp}(\Rvec_{ij},\{\rvec_k\}_{k\neq i,j})
         A^{\beta}_{pp} (\Rvec_{ij},\{\bs{r}_{k}\}_{k\neq i,j}),
\cr & =
     Z + Z(Z-1) \sum_{\alpha\beta} \bra A^{\alpha\dagger}_{pp} | A^{\beta}_{pp}  \ket 
            h_{pp}^{\alpha\beta}(\qvec)
\end{align}
where 
\be
  h_{pp}^{\alpha\beta}(\qvec)=\int d\rvec
                  \varphi_{pp}^{\alpha\dagger}(\rvec) e^{i\qvec\cdot\rvec} \varphi_{pp}^{\beta}(\rvec)
\ee
is a universal proton-proton function independent of the particular nucleus or its quantum state.
Averaging this result over the magnetic projection $M$ and utilizing Eq. (\ref{ave_contacts_def})
we finally get for $q \rightarrow \infty$
\be\label{CSR}
   CSR = Z+ \sum_{\alpha\beta} \frac{ 2 C^{\alpha\beta}_{pp}}{16\pi^2}h_{pp}^{\alpha\beta}(\qvec)
         - \rho_c^2(\qvec),
\ee
where $ \rho_c(\qvec) = \bra \Psi | \hat\rho_c(\qvec) | \Psi \ket $ 
is the nuclear charge distribution.
Equation (\ref{CSR}) relates the CSR to the proton-proton contact, as ultimately it 
should, since the CSR is a measure of charge fluctuations in the nucleus and on short
length scales these are completely dominated by two protons coming close together.
The CSR can be effectively probed experimentally in deep inelastic 
electron scattering experiments in which the virtual photon explores medium 
and short internucleon distances. 
In this regime the longitudinal nuclear structure function is sensitive to the pp SRCs
 which can modify the way the CSR approaches its model independent limit.
The connection between the 
CSR and the contact provides
a principle way to extract the proton-proton contacts from the available experimental 
electron scattering data.
We note that the CSR for point-like particles is closely related to a 
quantity called the static structure factor, which was previously shown to be
related to the single-channel s-wave contact for two-component fermions
in the zero-range case \cite{SSF1,SSF2}.

\section{Summary} 

Summing up, we have reviewed the derivation of Tan's relation for the one and two-body
momentum distributions and their generalization to nuclear physics. 
Though Tan's relations were originally experimentally established for cold atomic 
systems \cite{SteGaeDra10,SagDraPau12,ParStrKam05,WerTarCas09,KuhHuLiu10}, 
 due to their universality they are also of significance in nuclear physics
in which the energy scales are orders of
magnitude larger and length scales are many orders of magnitude smaller.  
We have seen that
the asymptotic $k\longrightarrow\infty$ tail of these distributions can be written as a
product of the contact matrix and a universal two-body function, highlighting the role
of few-body dynamics within a larger many-body system.

In general, the contact is a measure for the probability of finding a 
particle pair close to each other, it is thus a measure of the nuclear SRCs.
While mean-field calculations provide important information on
the nuclear shell structure, they do not constitute a full description of the structure of nuclei.
Especially, deviations are large for small, dense nuclear structures. 
It is long been known that 
the strong, short-range component of the nucleon-nucleon potential generates a high-momentum 
tail in the nucleon momentum distribution. We have demonstrated how this phenomena can
be described in terms of the contact.
Employing high-energy electromagnetic probes, short scale fluctuations in the nucleus can
be studied and information can thus be obtained regarding the different nuclear contacts.
Whereas in \cite{WeissPRC15,WeissPRL15,Hen15,WeissEPJA16,Ciofi16}
a connection was found between the 
different nuclear $np$ contacts and electromagnetic experimental rates, 
in this paper we have shown how 
short range charge fluctuations are connected to the pp contact. Thus different
experiments can help extract information on the different nuclear contacts which can in
turn help us better understand the details of the short range nuclear structure.


\begin{acknowledgements} The authors would like to thank Giuseppina Orlandini for useful
discussions regarding the Coulomb sum-rule.
This work was supported by the Pazy foundation.
\end{acknowledgements}


\begin{thebibliography}{}
  
\bibitem{Tan08}
  S. Tan, Ann. Phys. (N.Y.) {\bf  323}, 2952 (2008); Ann. Phys.
  (N.Y.) {\bf 323}, 2971 (2008); Ann. Phys. (N.Y.) {\bf 323}, 2987 (2008).
\bibitem{Bra12}
  E. Braaten, in BCS-BEC Crossover and the Unitary Fermi
  Gas, edited by W. Zwerger (Springer, New York, 2012).
\bibitem{SteGaeDra10}
  J. T. Stewart, J. P. Gaebler, T. E. Drake, and D. S. Jin, Phys.
  Rev. Lett. 104, 235301 (2010).
\bibitem{SagDraPau12}
  Y. Sagi, T. E. Drake, R. Paudel, and D. S. Jin, Phys. Rev.
  Lett. {\bf 109}, 220402 (2012).
\bibitem{ParStrKam05}
  G. B. Partridge, K. E. Strecker, R. I. Kamar, M. W. Jack,
  and R. G. Hulet, Phys. Rev. Lett. {\bf 95}, 020404 (2005).
\bibitem{WerTarCas09}
   F. Werner, L. Tarruel, and Y. Castin, Eur. Phys. J. B
  {\bf 68}, 401 (2009).
\bibitem{KuhHuLiu10}
  E. D. Kuhnle, H. Hu, X.-J. Liu, P. Dyke, M. Mark, P.
  D. Drummond, P. Hannaford, and C. J. Vale, Phys. Rev.
  Lett. {\bf 105}, 070402 (2010).
\bibitem{Lev51}
  J. S. Levinger, Phys. Rev. {\bf 84}, 43 (1951).
\bibitem{HenSci14}
  O. Hen et al. (CLAS Collaboration), Science {\bf 346}, 614 (2014).
\bibitem{ArrHigRos12}
  J. Arrington, D. Higinbotham, G. Rosner, M. Sargsian, 
  Prog. Part. Nucl. Phys. {\bf 67}, 898 (2012).
\bibitem{Wiringa14}
  R. B. Wiringa, R. Schiavilla, Steven C. Pieper, and J. Carlson
  Phys. Rev. C {\bf 89}, 024305 (2014).
\bibitem{Fomin12}%
  N. Fomin et al., 
  Phys. Rev. Lett. {\bf 108}, 092502 (2012).
\bibitem{WeissPRC15}
  R. Weiss, B. Bazak, and N. Barnea , Phys. Rev. C {\bf 92}, 054311 (2015).
\bibitem{WeissPRL15}
  R. Weiss, B. Bazak, and N. Barnea , Phys. Rev. Lett. {\bf 114}, 012501 (2015).
\bibitem{Hen15}  
  O. Hen, L. B. Weinstein, E. Piasetzky, G. A. Miller, M. M. Sargsian, and Y. Sagi,
  Phys. Rev. C {\bf 92}, 045205 (2015).
\bibitem{WeissEPJA16}
  R. Weiss, B. Bazak, and N. Barnea , Eur. Phys. J. A {\bf 52}, 92 (2016).
\bibitem{BraKanPla11} 
  E. Braaten, D. Kang, and L. Platter,
  Phys. Rev. Lett. {\bf 106}, 153005 (2011).
\bibitem{AlvCioMor08}%
  M. Alvioli, C. Ciofi degli Atti and H. Morita, 
  Phys. Rev. Lett. {\bf 100}, 162503 (2008).
\bibitem{WerCas12}%
  F. Werner and Y. Castin, 
  Phys. Rev. A {\bf 86}, 013626 (2012).
\bibitem{zerorange}%
  H. A. Bethe and R. Peierls, 
  Proc. Roy. Soc. {\bf 148}, 146 (1935).
\bibitem{CioSim96}%
  C. Ciofi degli Atti and S. Simula,
  Phys. Rev. C {\bf53}, 1689 (1996). 
\bibitem{AlvCio13}
 M. Alvioli, C. Ciofi degli Atti, L.P. Kaptari,
 C.B. Mezzetti and H. Morita, Phys. Rev. C {\bf 87}, 034603 (2013).
\bibitem{Ciofi15}
C. Ciofi degli Atti, Phys. Rep. {\bf 590}, 1 (2015).
\bibitem{Ciofi16}
  M. Alvioli, C. Ciofi degli Atti, and H. Morita,
  arXiv:1607.04103v1 [nucl-th] (2016).
\bibitem{TavTer92}%
  M. L. Terranova, D. A. De Lima and J. D. Pinheiro Filho,
  Europhys. Lett. {\bf 9} 523 (1989); 
  O. A. P. Tavares and M. L. Terranova, J. Phys. G {\bf 18}, 521 (1992).
\bibitem{SSF1}%
  H. Hu, X.-J. Liu and P. D. Drummond,
  Europhys. Lett. {\bf 91} 20005 (2010).
\bibitem{SSF2}%
  E. D. Kuhnle, H. Hu, X.-J. Liu, P. Dyke, M.Mark, P.D. Drummond,
  P. Hannaford, and C.J. Vale,
  Phys. Rev. Lett. {\bf 105} 070402 (2010).
\bibitem{Drell1958}
  S. D. Drell and C. L. Schwartz,  Phys. Rev. {\bf 112} 568 (1958).
\bibitem{McVoy1962}
  K. W. McVoy and L. Van Hove,  Phys. Rev. {\bf 125} 1034 (1962).
\bibitem{GiusySR}
  G. Orlandini and M. Traiui, Rep. Prog. Phys. {\bf 54}, 257 (1991).
\bibitem{nature-pwave-2016}
  C. Luciuk, S. Trotzky, S. Smale, Z. Yu, S. Zhang, and J.
  H. Thywissen, Nature Phys. {\bf 12}, 599 (2016).
\bibitem{3BodyTerm} 
  P.F.~Bedaque, H.-W.~Hammer, and U. van Kolck,
  Phys. Rev. Lett. {\bf 82} (1999) 463;
  Nucl. Phys. A {\bf 676} (2000) 357.
\bibitem{pionlesseft}
  P.F. Bedaque and U. van Kolck,
  Ann. Rev. Nucl. Part. Sci. {\bf 52} (2002) 339.
\bibitem{Thomas35}
  L. Thomas, Phys. Rev. {\bf 47}, 903 (1935).
\bibitem{Efi70}
  V. Efimov, Phys. Lett. B {\bf 33}, 563 (1970). 

%
\bibitem{Bao-Jun16}
  C. Bao-Jun, and Li. Bao-An, Phys. Rev. C {\bf 93}, 014619 (2016).


\end{thebibliography}
\end{document}